\documentclass[useAMS,usenatbib]{mn2e}

\usepackage{amssymb}
\usepackage{graphicx}
\title[CSFR: a theoretical approach]{Cosmic star formation rate: a theoretical approach}
\author[L. Vincoletto \& al. ]{L. Vincoletto$^{1}$\thanks{E-mail: vincoletto@oats.inaf.it}
F. Matteucci$^{1,2}$ F. Calura$^{3}$ L. Silva$^{2}$ G. Granato$^{2}$\\
$^{1}$Dipartimento di Fisica - Sezione di Astronomia - Universit\`{a} degli Studi di Trieste - Piazzale Europa, 1, I-34127 Trieste, Italy\\
$^{2}$INAF - Osservatorio Astronomico di Trieste, Via G.B. Tiepolo 11, I-34131 Trieste, Italy\\
$^{3}$INAF - Osservatorio Astronomico di Bologna, Via Ranzani 1, I-40127 Bologna, Italy\\}
\begin{document}
\date{}

\pagerange{\pageref{firstpage}--\pageref{lastpage}} \pubyear{2011}

\maketitle

\label{firstpage}

\begin{abstract}
The cosmic star formation rate (CSFR), namely the star formation rate in a unitary comoving volume of the Universe, is an important clue 
to investigate the history of the assembly and evolution of galaxies. Here, we develop a method to study the CSFR from 
a purely theoretical point of view. Starting from detailed models of chemical evolution, which best fit the properties of local galaxies, 
we obtain the histories of star formation of galaxies of different morphological types (ellipticals, spirals, irregulars).
These histories are then used to determine the luminosity functions of the same galaxies by means of a spectro-photometric code. 
We obtain the CSFR under different hypotheses about galaxy formation scenarios. First, we study the hypothesis of a pure luminosity evolution 
scenario, in which all galaxies are supposed to form at the same redshift and then evolve only in luminosity without any merging or interaction.
Then we consider scenarios in which the number density or the slope of the luminosity functions are assumed to vary with redshift.
After comparison of our results with the data available in literature, we conclude that a pure luminosity evolution does not provide a
good fit to the data, especially at very high redshift, although many uncertainties are still present in the data because of the unknown dust corrections
and assumed initial mass function. On the other hand, a variation in the number density of ellipticals and spirals as a function of redshift can provide 
a better fit to the observed CSFR. We critically discuss the possible scenarios for galaxy formation derived from this finding. We also explore cases 
of variable slope of the luminosity functions with redshift as well as variations of number density and slope at the same time. We cannot find any of those 
cases which can fit the data as well as the solely number density variation. Finally, we compute the evolution of the average cosmic metallicity in 
galaxies with redshift and show that in the pure luminosity evolution case the ellipticals 
dominate the metal production in the Universe, whereas in the case of number density evolution are the spirals the main producers of metals. 
\end{abstract}

\begin{keywords}
galaxies: evolution - galaxies: fundamental parameters - galaxies: photometry - galaxies: luminosity function - ISM: general - ISM: evolution.
\end{keywords}

\section{Introduction} 

The rate at which galaxies have formed stars throughout the whole cosmic time is a fundamental clue to investigate the history of the 
assembling and evolution of structures in the Universe.
The cosmic star formation rate (CSFR), defined as the comoving space density of the global star formation rate in a unitary
volume of the Universe, is not a  directly measurable quantity. It can be inferred only from the measurement of the luminosity density in 
different wavebands, which are then transformed into star formation rate by suitable calibrations.
Starting from its first determination made by \citet{a1} and thanks to deeper and deeper surveys it has been possible to extend the 
determination of the cosmic star formation rate up to $z\sim10$ \citep{a2,a3,a4,a5,a6,a61,a8}.
Constraining the CSFR at high redshift would be of paramount importance to put constraints in the galaxy formation scenario.
In particular, to decide between the \textit{Monolithic Collapse} (MC) and the \textit{Hierarchical Clustering} (HC), because the CSFR depends on 
both the star formation rate (SFR) in galaxies and on the evolution of the galaxy luminosity function (LF).
In the MC scenario, spheroids and bulges formed at high redshift (e.g. $z>2-3$) as the result of a violent burst of star formation, following the 
collapse of a gas cloud. After the development of a galactic wind quenching star formation, galaxies evolved passively until present time \citep{a9,a10}. 
Moreover, \citet{a11} introduced in this scenario the assumption that more massive spheroids had higher star formation efficiencies than  
less massive ones, and stopped to form stars earlier. It is common now to refer to this behaviour as ``downsizing in star formation''. 
These assumptions allow us to reproduce the majority of chemical and photometric properties of local ellipticals. 
On the other hand, in the HC scenario for galaxy formation, baryon assembly basically mirrors the hierarchical build up of dark matter structures.
Therefore, spheroids and bulges formed as a series of subsequential mergers among gas-rich galaxies or with galaxies that have already stopped their 
star formation.
In this scenario, galaxies form stars at lower rates than in the MC scenario, with more massive spheroids reaching their final mass at later 
times ($z\lesssim1.5$) \citep{a12,a13,a14,a15}.\\
In this paper, we calculate the evolution of the CSFR and the mean metallicity of the interstellar medium (ISM) in galaxies.  
This is done by means of detailed chemical evolution models for galaxies of different morphological types, i.e., ellipticals, spirals, and 
irregulars, which successfully reproduce the local properties of such galaxies, and from which we derive their star formation histories (SFH).
Star formation histories obtained with chemical evolution models are then used to determine the photometric evolution of the galaxies
(spectra and magnitudes), through a spectro-photometric code.
This allows us to compute the LF, that gives the number of galaxies in a unitary volume of the Universe in a luminosity bin.
In this work, we normalize our LFs to the local LFs derived by \citet{a16}, which are based on the \textit{``Second Southern Sky 
Redshift Survey'' ($SSRS2$)} using data from $5404$ galaxies at $z\le0.05$. These LFs are determined in the B band
for galaxies of different morphological type. Then we compute the evolution in luminosity of the galaxy at the break of the LF for each 
galactic type. This is done under different scenarios: first, we study the hypothesis of a pure luminosity evolution 
scenario (PLE). In this case, galaxies are supposed to form all at the same redshift and then evolve only in luminosity without any merging or 
interaction. In this case, we follow the approach of \citet{a17}. 
Then we define two new scenarios: one in which the number density of the LF is assumed to vary with redshift, and one in which the slope of the LF is
assumed to vary with redshift.
All the results have been compared with the available data found in literature.
Finally, we calculate the mean metallicity of the ISM for the PLE  and the number density evolution (NDE) scenario.
The paper is organized as follows: in \S~2 we describe the chemical and photometric models, in \S~3 we present our results and in \S~4 we 
draw some conclusions.

Throughout this paper the cosmology adopted will follow the ${\Lambda}CDM$ paradigm with ${\Omega}_{0}=0.3$,
${\Omega}_{\Lambda}=0.7$ and $h=0.65$.

\section{Chemical and Photometric evolution of galaxies} 

Chemical evolution models allow us to compute the evolution in time and space of several quantities such as the star formation, the production 
rate of chemical elements and the chemical abundances in the stars and in the gas, starting from the matter reprocessed by the stars and 
restored into the ISM through stellar winds and supernova explosions.
Starting from observational constraints such as the present day chemical abundances, it is possible to perform a detailed backward evolution of 
galaxies of different morphological type.
We consider three galactic types namely irregulars, spiral disks and spheroids which include both bulges and elliptical galaxies. 
This is motivated by the fact that ellipticals and bulges are likely to have a common origin, i.e. both are likely to have formed their stars on 
very short timescales and a long time ago \citep{a52}.
Detailed descriptions of the adopted models can be found in \citet{a18} for spirals, \citet{a19} for ellipticals and \citet{a20} for irregulars.

In our picture, spheroids form as the result of a sudden collapse of a gas cloud of primordial chemical composition. The gas is supposed  
to be accreted over a finite timescale, so in this respect we do not assume pure MC models. 
A galaxy can be modelled either as a one-zone object or a multi-zone object in which the galaxy is divided into several concentric shells independent of each-other. 
The spheroids are modelled as a one zone objects and they are supposed to suffer a strong initial burst of star formation, that stops as soon as the galactic 
wind develops; in this sense their evolution crucially depends on the time at which the galactic wind occurs. From that moment on the galaxy is supposed to evolve passively.

The chemical evolution of spiral disks is studied by using the one-infall model of \citet{a18}, developed to reproduce the
observational constraints of the Milky Way disk. This model belongs to the class of the so-called \textit{``infall models''}, where the disk is
thought as gradually forming by infall of primordial gas. Instantaneous mixing of gas is assumed, but the instantaneous recycling approximation is 
relaxed. The galactic disk is supposed to have formed on timescales increasing with the galactocentric distance, being larger at larger radii, in 
agreement with the dissipative models of galaxy formation of \citet{a21}. The gas accumulates faster in the inner than in the outer region, according 
to the so-called \textit{``inside-out''} scenario. The process of disk formation is much longer than the halo and the bulge formation (they should
form on timescales no longer than $1$~Gyr) with typical timescales varying from $\sim ~2$~Gyr in the inner disk, $\sim ~7$~Gyr in the solar region 
and up to $15-20$~Gyr in the outer disk. This mechanism is important to reproduce the observed abundance gradients (see \citealt{a18}, \citealt{a53}). 
The spiral disk is approximated by several independent rings, $2$ kpc wide, without exchange of matter between them. In this work the disk has been 
set to extend from $2$ to $16$~kpc.

Irregular galaxies are modelled as one zone objects and are assumed to assemble all their gas by means of a continuous infall of primordial gas and to form 
stars at a lower rate than the other morphological types. Their stellar populations appear to be mostly young, their metallicity is low and their 
gas content is large. All these features indicate that these galaxies are poorly evolved objects. Furthermore, irregular galaxies are
allowed to develop moderate galactic winds which do not halt the star formation as in the spheroids \citep{a20}.

\subsection{The chemical evolution models} 
The equation that describes the evolution of the \textit{i}-th element has the following form:

\begin{eqnarray}
&&{d G_{i}(t) \over d t} = -{\psi(t)}{X_{i}(t)} \nonumber \\
&& +\int^{M_{Bm}}_{M_{L}}{\psi}(t-{\tau}_{m})Q_{mi}(t-{\tau}_{m}){\phi}(m)\, dm \nonumber \\
&& +A\int^{M_{BM}}_{M_{Bm}}{\phi}(m)\ \nonumber \\
&& {\cdot} \Biggl{[}\int^{0.5}_{{\mu}_{min}}f({\mu}){\psi}(t-{\tau}_{m2})Q_{mi}(t-{\tau}_{m2}){\phi}(m)\, d{\mu}\Biggr{]} dm \nonumber \\
&& +(1-A)\int^{M_{BM}}_{M_{Bm}}{\psi}(t-{\tau}_{m})Q_{mi}(t-{\tau}_{m}){\phi}(m)\, dm \nonumber \\
&& +\int^{M_{U}}_{M_{BM}}{\psi}(t-{\tau}_{m})Q_{mi}(t-{\tau}_{m}){\phi}(m)\, dm \nonumber \\
&& + X_{Ai}A(t)-X_{i}W(t)
\,. \label{eq:chev}
\end{eqnarray}

Where $G_{i}(t)$ is the normalized fractional mass of gas within a galaxy in the form of the element $i$:
\begin{equation}
G_i(t)\;=\;\frac{M_g(t)\,X_i(t)}{M_{tot}} \, ,
\end{equation}
where $M_{tot}$ and $M_g(t)$ are the total galaxy mass and the mass of gas at time $t$, respectively.
\begin{equation}
X_{i}(t)\;=\;\frac{G_{i}(t)}{G(t)}
\end{equation}
represents the abundance in mass of the element $i$, the summation over all elements in the gas mixture being equal 
to unity. Thus, the quantity: 
\begin{equation}
G(t)= \frac{M_{g}(t)}{M_{tot}},
\end{equation}
is the total fractional mass of gas present in the galaxy at time $t$. In the case of spiral disks $M_{g}(t)$ and $M_{tot}$ are replaced by the 
corresponding surface densities $\sigma(r,t)$ and ${\sigma}(r,t_{G})$

The first term on the right hand side of eq. \ref{eq:chev}, including the star formation rate ${\psi}(t)$, represents the rate at which each element disappears from the 
ISM owing to the star formation. 

The second term is the rate at which each element is restored into the ISM by single stars with masses in the range
$[M_{L},~M_{Bm}]$, where $M_{L}$ is the minimum stellar mass contributing, at a given time \textit{t}, to chemical enrichment ($\simeq 0.8 M_{\odot}$) 
and $M_{Bm}$ is the minimum total binary mass of systems which can give rise to Type Ia SNe ($3$ $M_{\odot}$, \citealt{a22}).
The quantity $Q_{mi}(t-{\tau}_{m})$ indicates the fraction of mass restored by the stars in form of an element \textit{i} that can be produced or 
destroyed in stars or both. This is the so-called \textit{``production matrix''} \citep{a23}.

The third term represents the enrichment due to binaries which become Type Ia SNe, i.e. all the binary systems with total mass in the range between
$M_{Bm}$ and $M_{BM}$, with $M_{BM}$ being the maximum total mass of a binary system which can give rise to Type Ia SNe and is equal to $16$ $M_{\odot}$ ($8+8$ $M_{\odot}$).
The model adopted for the progenitors of the Type Ia SNe is the single degenerate one as in \citet{a22}. 
The parameter \textit{A} represents the fraction of binary stars giving rise to Type Ia SNe, and although its real value is unknown it is fixed in
order to reproduce the observed present time SN Ia rate. This parameter obviously depends on the adopted IMF and, in general, values between
$0.05$ and $0.09$ are adopted. This ensures to reproduce the actual SN Ia rate both in the Milky Way and in other galaxies.
It is worth noting that, in this term, the quantities ${\psi}$ and $Q_{mi}$ refer to the time ($t-{\tau}_{m2}$), where ${\tau}_{m2}$ indicates the 
lifetime of the secondary star of the binary system, which regulates the explosion timescale of the system. 
The coefficient ${\mu} = M_{2}/M_{B}$, is the ratio between the mass of the secondary component and the total mass of the binary system, while
$f({\mu})$ is the distribution function of this ratio. It is calculated on a statistical basis indicating that values close to $0.5$ are preferred.
Its analytical expression can be written as:
\begin{equation}
 f({\mu})=2^{(1+{\gamma})}(1+{\gamma}){\mu}^{\gamma},
\end{equation}
with ${\gamma}=2$ as a parameter \citep{a24} and ${\mu}_{\min}$ is the minimum mass fraction contributing to the SNIa rate at the time 
\textit{t}, and is given by:
\begin{equation}
 {\mu}_{min}=max\Biggl{[}\frac{M_{2}(t)}{M_{B}}, \frac{M_{2}-0.5{M_{B}}}{M_{B}}\Biggl{]}.
\end{equation}
The fourth term of eq. \ref{eq:chev} represents the enrichment due to stars in the range $[M_{Bm}, M_{BM}]$ which are single or, if binaries, do not produce
a SNIa event. All the stars in this range with mass larger than $8 M_{\odot}$ are assumed to explode as core collapse supernovae. 

The fifth term represents the contribution to the chemical enrichment coming from stars more
massive than $M_{BM}$ (supposed to explode like Type II and I$_{b/c}$ SNe). In all the models the upper mass limit contributing to chemical 
enrichment is assumed to be $100$ $M_{\odot}$. 

Finally, the last two terms are the rate of accretion of matter with primordial abundances $X_{A}$, and the outflow rate for the element \textit{i},
respectively.

\subsubsection{The Star Formation Rate} \label{sec:sfr}
The main feature characterizing a particular morphological galactic type is represented by the prescription adopted for its star formation history.
In the case of elliptical galaxies the SFR ${\psi}$ has the simple form given by:
\begin{equation} \label{sfe}
 {\psi}~=~{\nu}G(t),
\end{equation}
where ${\nu}$ is the star formation efficiency, namely the inverse of the typical time scale for star formation, and assumed to be
$15$~Gyr$^{-1}$.

In the case of irregular galaxies, a continuous star formation rate is assumed as in eq. \ref{sfe}, but characterized by a lower efficiency than 
the one adopted for ellipticals (0.1 Gyr$^{-1}$ in this case). 

In the case of spiral galaxies the formulation is the following (see \citealt{a18} and \citealt{a54}):
\begin{equation}
 {\psi}(r,t)={\nu}{\cdot}\Bigg{[}{\frac{\sigma(r,t)}{{\sigma}({r_{\odot}},t)}\Bigg{]}^{2(k-1)}}{\cdot}\Bigg{[}{\frac{\sigma(r,t_{G})}{{\sigma}(r,t)}\Bigg{]}^{(k-1)}}{\cdot}G^{k}(r,t),
\end{equation}
where ${\nu}$ is set in this case equal to $1~$~Gyr$^{-1}$, ${\sigma}(r,t)$ is the total surface mass density at a given radius and 
at a given time, ${\sigma}(r_{\odot},t)$ is the total surface mass density at the position of the Sun (set equal to $8.0$~Kpc). 
Finally ${\sigma}(r,t_{G})$ is the surface mass density at present time. 
The exponent \textit{k} is assumed equal to 1.4, as suggested by \citet{a55}.

All these models are tuned to reproduce at best the main observed properties of local galaxies of different morphological type.
In Table \ref{tab:car} we show the main characteristics of the galaxies of different morphological type that we use in this work.
It is important to stress that in our models, we consider only one galaxy per morphological type and take it as representative 
of the whole population. For the spiral galaxy we take the average SF and metallicity of the disk.
\begin{table}
\centering

\begin{tabular}{@{}llccc@{}}
\hline
$Morphological$& $M_{tot}$& $\tau_{inf}$& $\nu$& $R_{eff}$\\
$Type$ &($M_{\odot}$) &($Gyr$) &($Gyr^{-1}$) &($Kpc$)  \\     
\hline
Ellipticals& $10^{11}$& ${0.3}$& ${15}$& ${3.0}$ \\
Spirals&     $4{\times}10^{10}$& ${7.0}$& ${1.0}$& ${3.5}$ \\
Irregulars& ${10^{10}}$& ${0.5}$& ${0.1}$& ${1.0}$ \\ \hline
\end{tabular}
\begin{footnotesize}
\caption{Model parameters for the galaxies reproduced in this work. $M_{tot}$ is the baryonic mass, ${R_{eff}}$ is the
         effective radius, ${\tau}_{inf}$ is the infall timescale and ${\nu}$ is the star formation efficiency.
         For spirals we indicate the average infall time scale for the disk.} \label{tab:car} 
\end{footnotesize}
\end{table}

\begin{figure}
\centering
\includegraphics[height=20 pc, width=20 pc]{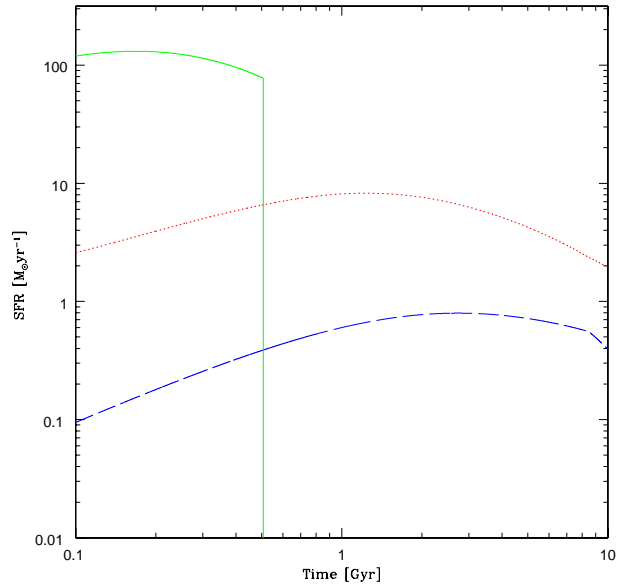}
\begin{footnotesize}
\caption{Star formation rates as a function of time, for a typical elliptical of $10^{11}~M_{\odot}$ (green solid line), spiral of 
        $\sim 4\times10^{10}~M_{\odot}$ (red dotted line) and irregular of $10^{10}~M_{\odot}$ (blue dashed line) galaxy.} \label{fig:sfh} 
\end{footnotesize}
\end{figure}

In Figure \ref{fig:sfh} we show the histories of star formation of the galaxies of different morphological type described above, for a \citet{a25} 
IMF. It is worth noting that for a spiral like the Milky Way an IMF steeper than the one used is usually adopted. Here we assume \citet{a25}
IMF for all galaxies because the CSFR derived observationally is obtained by means of this IMF.
As it is clear from the Figure a typical elliptical ($10^{11}~M_{\odot}$ of luminous mass) experiences a high burst of star formation lasting 
for about $0.6$~Gyr with a maximum peak of more than $\sim 100~M_{\odot}yr^{-1}$. After such a period, the star formation ceases abruptly owing to 
the onset of the galactic wind. The spiral disk is characterized by a continuous SFR with a large peak around $1\sim2$~Gyr of less than 
$\sim 10~M_{\odot}yr^{-1}$ and a present time value of $\sim2~M_{\odot}yr^{-1}$ (for the Milky Way). Finally, the irregular galaxy forms stars at a 
rate smaller than the previous morphological types, with an increasing SFR that reaches a maximum of less then $1~M_{\odot}yr^{-1}$, with a 
smaller present value, due to the onset of the galactic wind.

\subsection{The photometric model}
To study the evolution of the photometric properties of galaxies, we use the spectro-photometric code \textsc{grasil} \citep{a26}.
Its starting point is the output of a chemical evolution code that follows the time evolution of SFR, metallicity (Z) and gas fraction.
The building blocks of galaxy photometric models are the library of isochrones and the stellar atmospheric models. \textsc{grasil} is based on the Padova 
stellar models \citep{a28}, with a major difference consisting in the computation of the effects of dusty envelops around AGB stars, and 
on the stellar atmospheric models by \citet{a27}. The single stellar populations (SSPs) span a range from $1~Myr$ to $18~Gyr$ for what concerns ages, 
and $Z=0.004,~0.008,~0.02,~0.05,~0.1$ for the metallicities (keeping the relative proportion of the metals equal to the solar partition).
At this point, the spectral synthesis is performed summing up the spectra of each stellar generation provided by the SSP of the appropriate age 
and metallicity, weighted by the $SFR$ (${\psi}$) at the time of the star birth. 
The flux at a given wavelength is defined as:
\begin{center}
\begin{equation}
 F_{\lambda}(t_{G})=\int_0^{t_{G}} {\psi}(t){\times}{SSP_{\lambda}}(t_{G}-t,Z(t))\,dt, \label{eq:ssp} 
\end{equation} 
\end{center}
where $F_{\lambda}(t_{G})$ is the integrated monochromatic flux at the present time \textit{$t_{G}$} and $SSP_{\lambda}(t_{G}-t,Z(t))$ is the 
integrated monochromatic flux of an SSP at age \textit{t} and metallicity Z.    
If the effects of dust are neglected (as in our case) the spectral energy distribution (SED) is simply given summing up the spectra of all stars.

\subsection{The luminosity evolution of galaxies}\label{sec:leg}
To derive the CSFR, it is of fundamental importance the knowledge of the relative distribution of 
galaxies, of any morphological type, as a function of redshift. This can be done through the LF, which gives the distribution of galaxies per unit 
volume in the luminosity interval $[L,~L+dL]$.
The LF is well reproduced by the \citet{a29} function:
\begin{equation}
 {\Phi}(L)\frac{dL}{{L}^{*}}={\phi}^{*}{\Biggl{(}}\frac{L}{{L}^{*}}{\Biggr{)}}^{-\alpha}e^{(-L/{L}^{*})}\frac{dL}{{L}^{*}}, \label{eq:sche}
\end{equation}
where ${\phi}^{*}$ is the number of galaxies per unit volume; ${L}^{*}$ is the characteristic luminosity which separates bright sources from faint 
sources and ${\alpha}$ is the slope of the luminosity function. 
Different determinations of the local LF exist (see \citealt{a30} for a review) and they show several differences, in the
sense that the parameters of the Schechter function are not fully constrained by observations. The luminosity function can be measured in several 
bands, obviously depending on the redshift regime under investigation. 
In this work, we make use of the determination of the local B-band luminosity function of \citet{a16}, that has the 
characteristic of considering separately the different galactic morphological types. For each morphological type the parameters of the LF
are summarized in table \ref{tab:marzke}.
\begin{table}
\centering
\begin{tabular}{@{}lccc@{}}
\hline
$Morphological$& $M^{*}$& $\alpha$& $\phi^{*}$\\
$Type$\\      
\hline
Ellipticals& $-19.37$&  $-1.00$& $4.4$ \\
Spirals&     $-19.43$& $-1.11$& $8.0$ \\
Irregulars& $-19.78$& $-1.81$& $0.2$ \\ \hline
\end{tabular}
\begin{footnotesize}
\caption{Parameters of the LF as defined in \citet{a16}. Here $M^{*}$ refers to the magnitude of the galaxy at the break, $\phi^{*}$
refers to the normalization of the LF (expressed in $Mpc^{-3}$) and $\alpha$ is the slope of the LF.} \label{tab:marzke} 
\end{footnotesize}
\end{table}

Once we have the LF, we can compute the \textit{luminosity density (LD)}, which is what is 
really observed. The luminosity density, in a given band, is defined as the integrated light radiated per unit volume from the entire galaxy 
population. It stems from the integral over all luminosities of the observed luminosity function:
\begin{equation}
 {\rho}_{L}=\int{\Phi}(L){\Biggl{(}}\frac{L}{{L}^{*}}{\Biggr{)}}\,dL. \label{eq:ld}
\end{equation}
At $z=0$, the LDs for the single galaxy types are given by the above integral of the present time luminosity functions of \citet{a16}.
At redshifts other than zero, for each morphological type we consider the luminosity evolution obtained with the spectrophotometric code.

\section{Results} 
As already pointed out, the CSFR is not a  directly measurable quantity. It can be inferred only from the measurement of the luminosity density in 
different wavebands and then converted into SF by means of suitable calibrations. Its first determination was made by \citet{a1} (at $2800~${\AA}, $4400~${\AA}, 
and $1~{\mu}m$), followed in the subsequent years by \citet{a31} (using the $[OII]$ line) and \citet{a32} (in the $UV$). 
These studies constrained the CSFR from $z=0$ to $z=1$, establishing that it is steadily increasing in that range.
The CSFR seems to peak around $z\sim2-3$ and then decrease at higher redshifts, where there is not a well defined trend. For example \citet{a4} on the
basis of optical and UV data and \citet{a3} on the basis of a compilation of data taken in different bands, found that the CSFR from roughly 
$z\sim3$ to $z\sim9$ is steadily decreasing. On the other hand, \citet{a6}, on the basis of the gamma ray burst (GRB) data from \textit{Swift} found a CSFR that is 
roughly constant from $z\sim4$ to $z\sim8$. 

\subsection{The cosmic star formation rate}
By means of the star formation histories of galaxies of different morphological type and their photometric evolution we can calculate the CSFR 
according to the method adopted by \citet{a17}:
\begin{equation} 
 \dot{\rho}_*(z)=\sum_{i}{\rho}_{Bi}(z)\Bigg{(}{\frac{M}{L}}\Bigg{)}_{Bi}(z){\psi}_{i}(z),  \label{eq:csfr}
\end{equation} 
where $\dot{\rho}_*(z)$ is the SFR density, ${\rho}_{Bi}$ is the B-Band luminosity density, $\Big{(}{\frac{M}{L}}\Big{)}_{Bi}$ is the B-band 
mass-to-light ratio and ${\psi}_{i}$ (expressed in $yr^{-1}$) is the star formation rate per unit mass of the \textit{i}-th morphological type.

In this work, we make use of the B-band luminosity, because it is a good star formation tracer since hot massive stars emit in this band.

In principle, the CSFR, could be computed simply by summing the SFR of each galaxy type multiplied by the corresponding number density. 
However, this simple approach would not allow us to explore in detail the case in which the slope of the LF varies with redshift 
(see section \ref{sec:alpha}). We tested that the two approaches lead to the same result.

\subsubsection{The Pure Luminosity Evolution Scenario}

Here we describe the determination of the CSFR in the framework of a PLE scenario. 
With pure luminosity evolution, we mean that galaxies of all morphological types are supposed to form at high redshift and then 
to evolve only in luminosity and not in number. We also assume that the slope of the LF is constant with redshift.
In this context, galaxies are assumed to be isolated, i.e. the effects of mergers or interactions are irrelevant at all redshifts, and it is
assumed that they start forming stars all at the same time.

In this simple approach, the differences between the various morphological types should be ascribed only to their different star formation histories. 
Early-type galaxies (i.e. ellipticals, bulges and S0) are dominated by old stellar populations so they must have formed their stars with high efficiency
and in a short period, while late-type galaxies (i.e. spirals and irregulars) should have formed stars for the whole Hubble time, as seen in Fig. \ref{fig:sfh}. 

\begin{figure} 
\centering
\includegraphics[height=20 pc, width=20 pc]{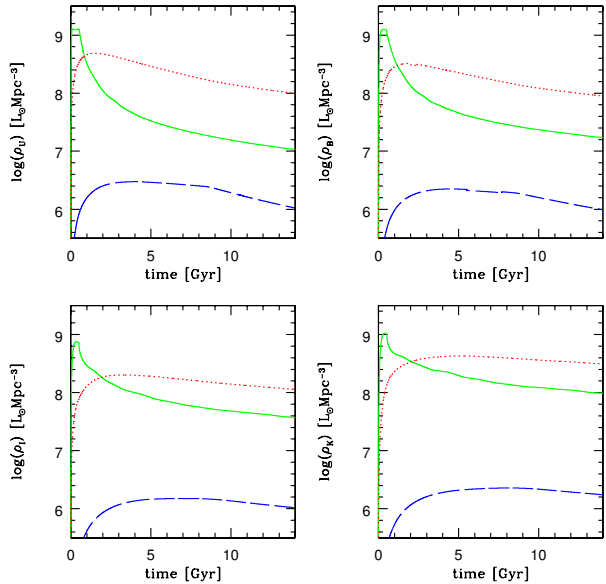}
\begin{footnotesize}
\caption{Prediction of the luminosity density evolution for galaxies of different types in U (top left), B (top right), I (bottom left) and K 
         (bottom right) band respectively, in the case of pure luminosity evolution. Here the x-axis represents the elapsed time since z=10.
         Solid green line refers to ellipticals, red dotted line to spirals and long dashed blue line to irregulars.}\label{fig:ldens}  
\end{footnotesize}
\end{figure}

In Figure \ref{fig:ldens} we can see the predicted evolution of the luminosity density for ellipticals, spirals and irregulars, in the
U (centered at $3650$~{\AA}), B (centered at $4450$~{\AA}), I (centered at $8060$~{\AA}) and K (centered at $21900$~{\AA}) band. These
evolutions are computed by adopting the SFRs of Fig. \ref{fig:sfh} and constant  slope and number density in the luminosity function.

From the Figure it is clear that, at early times in every band, the luminosity density is dominated by the light emitted by elliptical
galaxies. This is obviously due to the fact that in the monolithic scenario they all suffer a strong initial burst of star formation 
which lasts for $<1~Gyr$ and then evolve passively. 
At this time a galactic wind develops and star formation stops, so their luminosity in the U and B bands,
where young, newborn stars emit, begins rapidly to decrease.
On the other hand, spirals form stars continuously and the decrease in luminosity in the U and B band should be ascribed
mostly to the consumption of their gas reservoir. Finally, also irregulars form stars for the entire cosmic time but their 
luminosity density is lower than that of the other morphological types due to their low star formation. Their luminosity density is also decreased
by the onset of galactic winds at late times.

The I and K bands are dominated by the light emitted by less massive, long living stars. Also in this case, ellipticals are the main 
sources at early times, while spirals become dominant for cosmic time $>2~Gyr$. However, the difference in the values reached by ellipticals and spirals is
less pronounced than in U and B bands. The luminosity density of spirals, after an initial phase of constant increase lasting $\sim3~Gyr$ is 
observed to decrease slowly in the I band and to be quite constant in the K band. 
Irregulars, also in this case have values lower than ellipticals and spirals which tend to decrease very slowly through the whole cosmic time.

The PLE model belongs to the category of the so-called \textit{``backward evolution''} models \citep{a33}; these models start from a well known 
and well constrained description of the local Universe to reconstruct the evolution of galaxies back to their formation. 
 
\begin{figure}
 \centering
 \includegraphics[height=20 pc, width=20 pc]{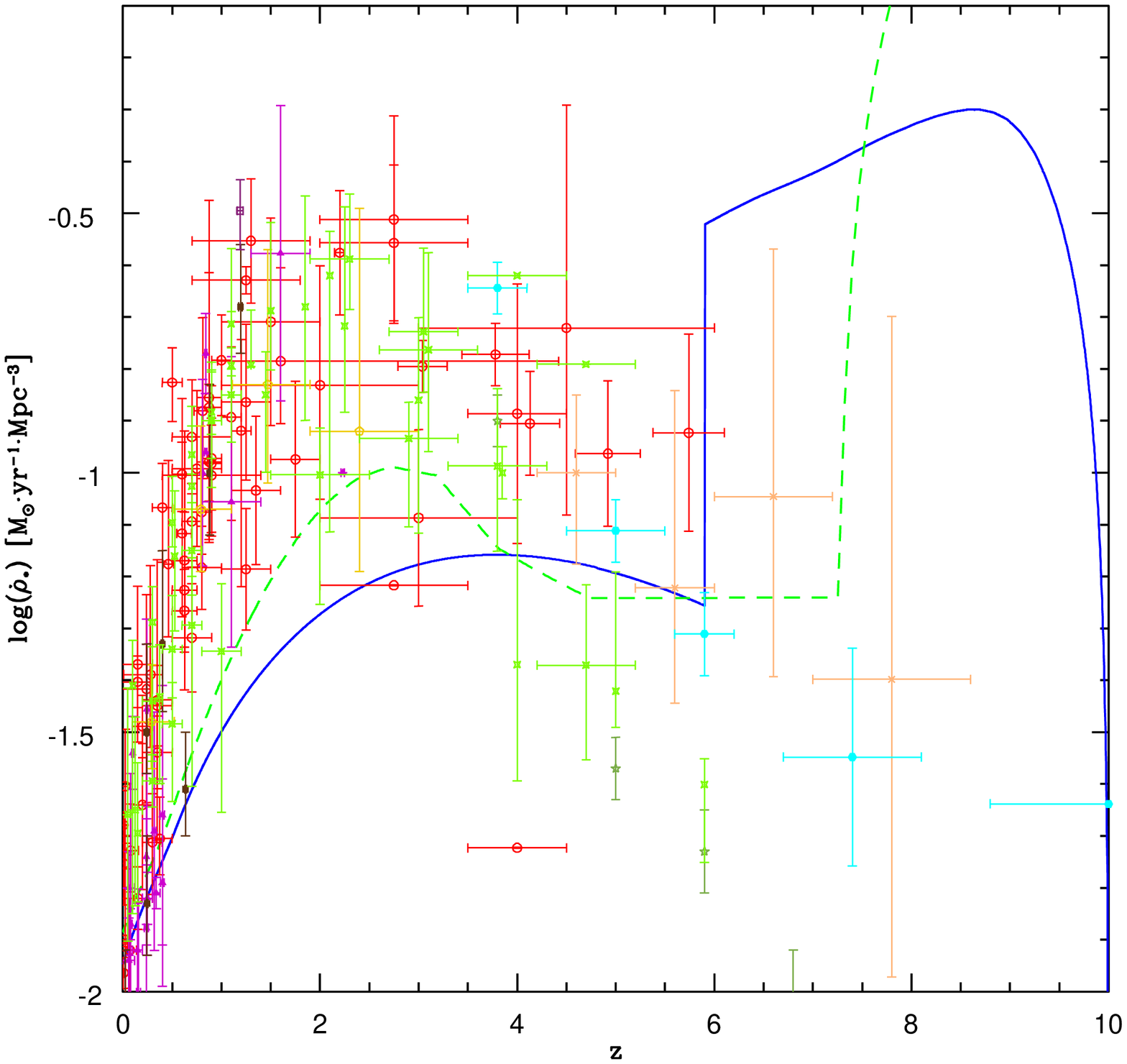}
 \caption{Cosmic star formation history in the case of pure luminosity evolution. Blue solid line, model described in this work; green dashed line 
          model from \citet{a17}. Points are taken from: \citet{a3} red open circles, \citet{a59} brown squares, \citet{a63} purple open squares,
          \citet{a5} cyan open hexagons, \citet{a62} yellow pentagons, \citet{a6} orange crosses, \citet{a60} magenta triangles, \citet{a61} green open stars,
          \citet{a64} light green crosses.}\label{fig:csfr1}
\end{figure}

In our model, all the galaxies are supposed to form at $z=10$. This choice is motivated by the fact that more and more objects are revealed at $z>5$
(\citealt{a34},\citealt{a35}), with evidences of a $z\sim10$ detection \citep{a7}.
The predicted CSFR is shown in Figure \ref{fig:csfr1}; for comparison we plot in the same Figure also the CSFR as obtained by \citet{a17} who adopted very 
similar assumptions and star formation histories.

As it can be seen, both models show some discrepancies when compared with the observational data. In our case, the discrepancy begins 
at $z\sim0.5$ and it is particularly evident in the redshift interval running from $\sim0.5$ to $\sim5$. Here we underpredict the observed values 
by a factor of $\sim3$. This is the redshift interval in which only spirals and irregulars contribute to the cosmic star formation
history. 
Also the Calura \& Matteucci CSFR suffers of the same problem when compared with data.
On the other side, at high redshift, our model seems to overpredict the data, as well as that of Calura \& Matteucci. The high observed peak is to ascribe to 
ellipticals which, in order to reproduce the local observational features, (like the increase of the $[Mg/Fe]$ ratio with galactic mass) 
have to form stars in a short time interval (lasting less than $1~Gyr$, \citealt{a11}). This implies that the SFR of these objects must lie 
in the range $\sim100-1000~M_{\odot}yr^{-1}$. At $z\sim6$ ellipticals stop to form stars and the CSFR decreases abruptly.
The discrepancies with the model of \citet{a17}  can be explained through the differences in the parameters adopted in the chemical evolution 
models and in the different photometric code used. In fact, in \citet{a17}, the photometric code of \citet{a56} was adopted.
Moreover, the chemical model for the ellipticals adopted by \citet{a17} is a closed box one, whereas our model includes gas infall.

\subsubsection{The Number Density Evolution Scenario (NDE)} \label{sec:nde}
In this section, we describe a different approach for the calculation of the CSFR. In what follows we introduce a scenario in 
which we let the number density of galaxies to evolve with redshift while the SFHs of galaxies are the same of Fig. \ref{fig:sfh}.

The main consequence of relaxing the hypothesis of the constancy of the number density of galaxies is that we consider galaxy interactions.
It has been claimed that giant ellipticals could form, as the result of the occurrence of 1 to 3 major dry-mergers during galactic lifetime
(\citealt{a37} and reference therein) and this view is in agreement with the observations.
Moreover it has been observed that the fraction of dry-mergers increases with redshift from $z\sim1.2$ to $z=0$ \citep{a38}.

To include the number density variation we modify the expression of the luminosity function as defined in eq. \ref{eq:sche}, letting the 
parameter $\phi^{*}$ (i.e. the number density of galaxies) to vary with the redshift. The new formulation of the luminosity function is:
\begin{equation}
 {\Phi}(L)\frac{dL}{{L}^{*}}={{\phi}_{0}{\cdot}{(1+z)}^{\beta}}{\Biggl{(}}\frac{L}{{L}^{*}}{\Biggr{)}}^{-\alpha}e^{(-L/{L}^{*})}\frac{dL}{{L}^{*}}, \label{eq:sche2}
\end{equation}
where $\phi_{0}=\phi^{*}(z=0)$, i.e. the present time number density of galaxies. This model is constrained in order to reproduce the number 
densities observed at $z=0$ by \citet{a16}. 

We let the parameter $\beta$ to run from $-6$ to $+6$ in steps of $0.2$. The choice of this interval has been made in order to 
explore a set of values as wide as possible without giving us meaningless results and, at the same time, letting us to test cases of extreme number 
density variation.
For each  value of $\beta$, we calculate the luminosity density using the modified version of the Schechter function. Then, using the star formation
histories depicted in section \ref{sec:sfr}, we determine the star formation rate density, using the equation \ref{eq:csfr}.

We perform a careful check of the contribution to the total CSFR from the different morphological types, and on its basis
we decide to let only spirals and ellipticals to evolve. Therefore, we consider the number density of irregulars constant through the cosmic time.
The reason resides in their contribution to the total CSFR, that is, at any redshift, two orders of magnitude lower than the one of ellipticals and spirals
and therefore negligible. Our best values of the $\beta$ parameter for each morphological type are:
\begin{itemize}
 \item Ellipticals: $\beta=-0.8$;
 \item spirals: $\beta=1$;
 \item irregulars: $\beta=0$.
\end{itemize}

\begin{figure}
\centering
\includegraphics[height=20 pc, width=20 pc]{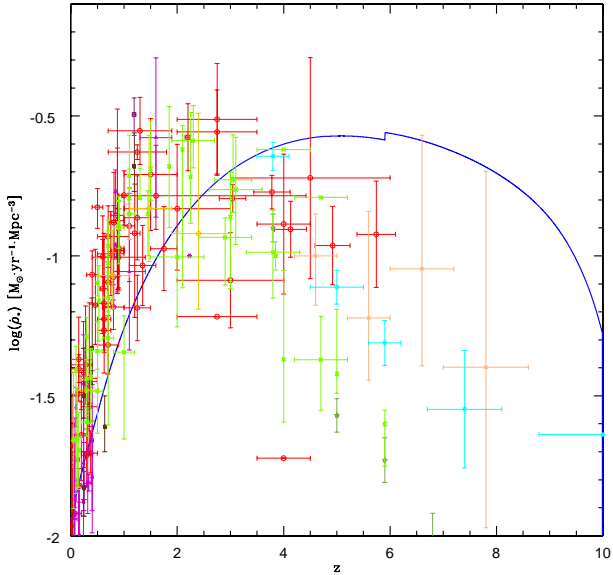}
\caption{The cosmic star formation history in the number density evolution scenario. Blue line refers to model results. Data points are the same 
         described in Figure \ref{fig:csfr1}.}\label{fig:csfr3}
\end{figure}

This means that the number density of ellipticals is supposed to increase from $z=10$ to $z=0$ (with $\sim15\%$ of ellipticals already in 
place at $z=10$). The opposite behaviour is predicted for spirals, which are supposed to decrease with decreasing redshift.
Under these assumptions, there is a possible interpretation in terms of galaxy formation. For example, the predicted number density evolution  
could mean that a large fraction of elliptical galaxies can form thanks to dry mergers of spirals, as proposed by \citet{a39} and \citet{a40}.
Since we are considering only dry-mergers it means that the merging process quenches the star formation in the progenitors and that
the residual gas is lost. From our model we can see that the number density of spirals decreases of roughly $60\%$ from $z\sim2$ to $z\sim0$, and of 
$50\%$ from $z\sim1$ to $z=0$. This is in good agreement with \citet{a41} who found a decrease of $50\%$ of Milky Way siblings from $z\sim1$ to $z=0$, indicating
this percentage as a lower limit. 
In Figure \ref{fig:csfr3}  we show our best model for the CSFR in the case of number density evolution. 
From the plot we can see that the peak observed in the PLE scenario completely disappears. The model seems to better reproduce the high redshift trend if the error bars are taken into account.
Also in the redshift interval between $0$ and $\sim5$ the agreement is improved although the curve is still slightly lower than the data, but this is a
problem common to all the other models of CSFR, as we will see later. 
Also in this case we underpredict the decrease of the CSFR between $z\sim2$ and $z=0$. This, in principle, can be due to the lack in our models of strong starburst galaxies
at low and intermediate redshift (see e.g. \citealt{a70}), even if as claimed by \citet{a71} starbursts seem to account only for the $10\%$ of the SFR density at $z\sim2$.
On the other hand, our predicted slope of the CSFR in this redshift range good as well as for the model of \cite{a65}. These authors using cosmological smoothed particle 
hydrodynamics simulations, explain this drop with a decline in the cold-accretion rate density on to haloes and with AGN feedback that decreases the contribution to the 
CSFR of the gas accreted in the hot mode.

\begin{figure}
\centering
\includegraphics[height=20 pc, width=20 pc]{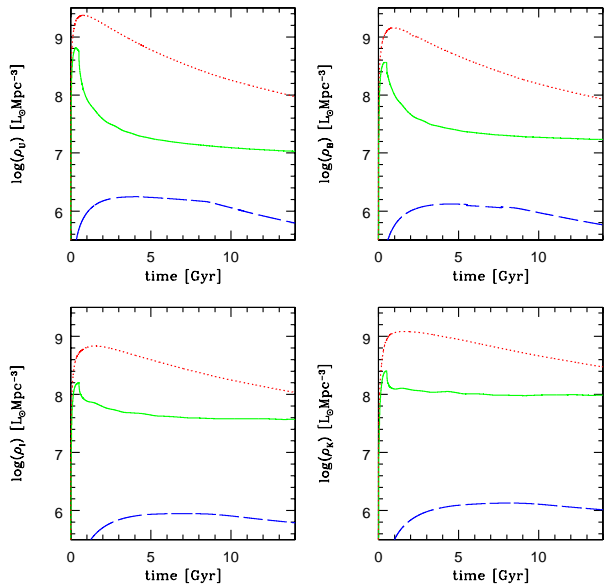}
\begin{footnotesize}
\caption{Prediction of the luminosity density evolution for galaxies of different types in U (top left), B (top right), I (bottom left) and K 
        (bottom right) band respectively in the case of number density evolution (best model). Here the x-axis represents the elapsed time 
        since z=10. Solid green line refers to ellipticals, red dotted line to spirals and long dashed blue line to irregulars.}\label{fig:ldens_nde}  
\end{footnotesize}
\end{figure}

In Figure \ref{fig:ldens_nde} we show the predicted evolution of the luminosity density, for our best model, in the case of number density 
evolution. Comparing these predictions with the ones in Figure \ref{fig:ldens}, it is possible to see that in this case spirals dominate the
luminosity density in every band and at any epoch, at variance with the PLE case.

It must be said that the variation of the number density of galaxies of different morphological types is still matter of debate,
since there is, in general, no concordance among various authors. For example, \citet{a42} found that, parametrizing the comoving number
density of $E/S0$ as proportional to $(1+z)^{\gamma}$, the favoured value for $\gamma$ is $-0.8\pm1.7$, in perfect agreement with our value.
This analysis is valid in the redshift range $0.3<z<0.8$, and the large error is due to the limited redshift range of their sample.
\citet{a43}, using data from the ESS (ESO-Sculptor Survey), found a number density evolution of late-type 
galaxies following the same parametrization, with $\gamma=2\pm1$. On the other hand, for example \citet{a44} found at $z<1$ an evolution of 
large disk structures (with a bulge-over-total ratio $B/T>0.5$) following a $(1+z)^{\gamma}$ relation, with $\gamma=\pm0.5$, lower than
the one found in our model.

Even more challenging is to asses the evolution at high redshift, since very deep surveys are required and the correspondence between actual galaxies
and their high $z$ counterparts is not firmly established. This causes a loss of information about the morphology of the various objects giving 
controversial results. Moreover, the uncertain dust corrections and the adopted IMF \citep{a25} used to derive the CSFR, could 
underestimate the actual CSFR. Therefore, we do not completely exclude that the CSFR could be constant or increasing at high z.

\subsubsection{The Variation of $\alpha$} \label{sec:alpha}
In this section we describe a modification of the Schechter function in which we consider the evolution with redshift of the slope of the LF for galaxies of different morphological type. 
We start again from the LF as defined in sec. \ref{sec:leg}, and assume that $\alpha={\alpha_{0}}{(1+z)}^{\beta}$, where $\alpha_{0}=\alpha(z=0)$ 
is the slope of the LF at present time and $\beta$ is a free parameter. Therefore, the LF assumes the form:

\begin{equation}
 {\Phi}(L)\frac{dL}{{L}^{*}}={{\phi}^{*}{\cdot}}{\Biggl{(}}\frac{L}{{L}^{*}}{\Biggr{)}}^{[{{-\alpha}_{0}}{(1+z)}^{\beta}]}e^{(-L/{L}^{*})}\frac{dL}{{L}^{*}}. \label{eq:sche3}
\end{equation}

This means that at any redshift we change the slope of the LF. 
As in the case of the number density evolution, we let $\beta$ to run from $-6$ to $+6$ in steps of $0.2$, and in no case we have found an improvement in fitting the data relative to the pure 
luminosity or number density variation scenarios. From the observational point of view, there are some suggestions of variations of the slope of the LF but referred to the whole population of galaxies, 
without differentiating for morphological types. For example, \citet{a66} found that $\alpha=-1.1$ locally and increases up to $\sim -1.7$ at high redshift (from z=3 to 6). From our numerical experiments 
this case corresponds to $\beta \sim 0.3$. We can exclude all the cases with $\beta>0.4$ and those with $\beta < -0.2$ since they predict values of $\alpha$ which are very much outside the observed range, 
and also a CSFR very much at variance with observations.

\begin{figure}
 \centering
 \includegraphics[height=20 pc, width=20 pc]{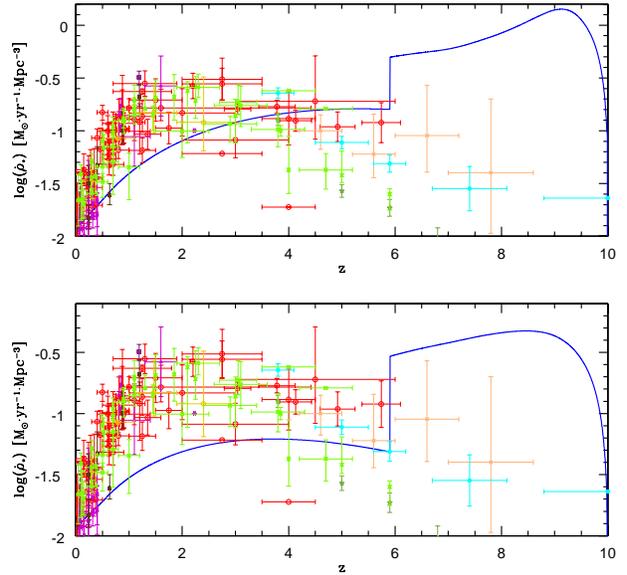}
 \caption{The CSFH in the case of variation of the LF slope $\alpha$. Here we show the cases corresponding to $\beta=0.4$ (top panel) and $\beta=-0.2$ (bottom panel). Blue lines indicate the results of the model.
          Data points are the same described in Figure \ref{fig:csfr1}. Note the different y-axis scale adopted for clarity.}\label{fig:alfa}
\end{figure}

\begin{figure}
 \centering
 \includegraphics[height=20 pc, width=20 pc]{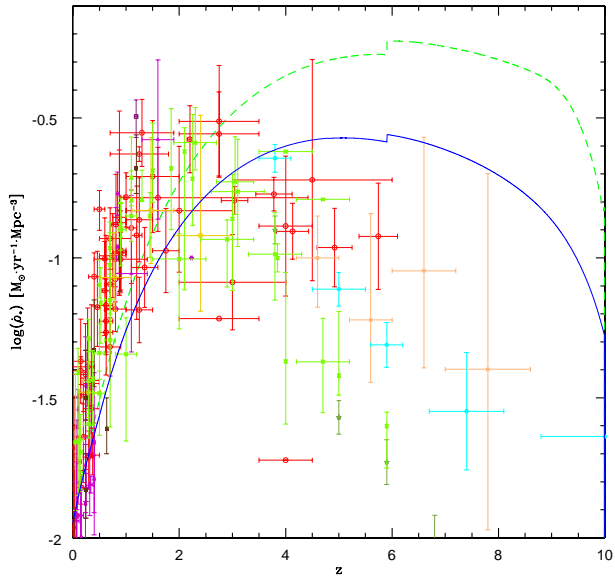}
 \caption{Comparison between our best model with NDE (blue solid line) and a model with both NDE (the same as the best model) and $\alpha$ evolution (green dashed line). 
          Data points are the same described in Figure \ref{fig:csfr1}.}\label{fig:alfa_bm}
\end{figure}

In Figure \ref{fig:alfa} we show the case $\beta=0.4$ (top panel) which is similar to the suggested observational variation \citep{a66}, and the case $\beta=-0.2$ (lower panel). 
Therefore, although we do not exclude a variation of the slope of the LFs of galaxies, we can conclude that the cases of $\alpha$ variations are providing a fit to the CSFR which is 
worse than the case in which there is number density variation, as shown in Figure \ref{fig:csfr3}. Finally, we run several models by varying both $\phi^{*}$ and $\alpha$ and we did 
not find any combination of parameters which can fit the observed CSFR better than our best case with  $\phi^{*}$ only variation. In Figure \ref{fig:alfa_bm} we show a comparison 
between a model with the same variation of $\phi^{*}$ as our best model (see section \ref{sec:nde}) and a model with the same $\phi^{*}$ variation plus a variation of $\alpha$, obtained with 
$\beta=0.3$ for ellipticals and   $\beta=0.2$ for spirals and irregulars. As one can see, the agreement between the model with variable $\alpha$ and $\phi^{*}$ does not fit 
the data as well as our best model.

\subsubsection{Comparison with other determinations}
\begin{figure}
\centering
\includegraphics[height=20 pc, width=20 pc]{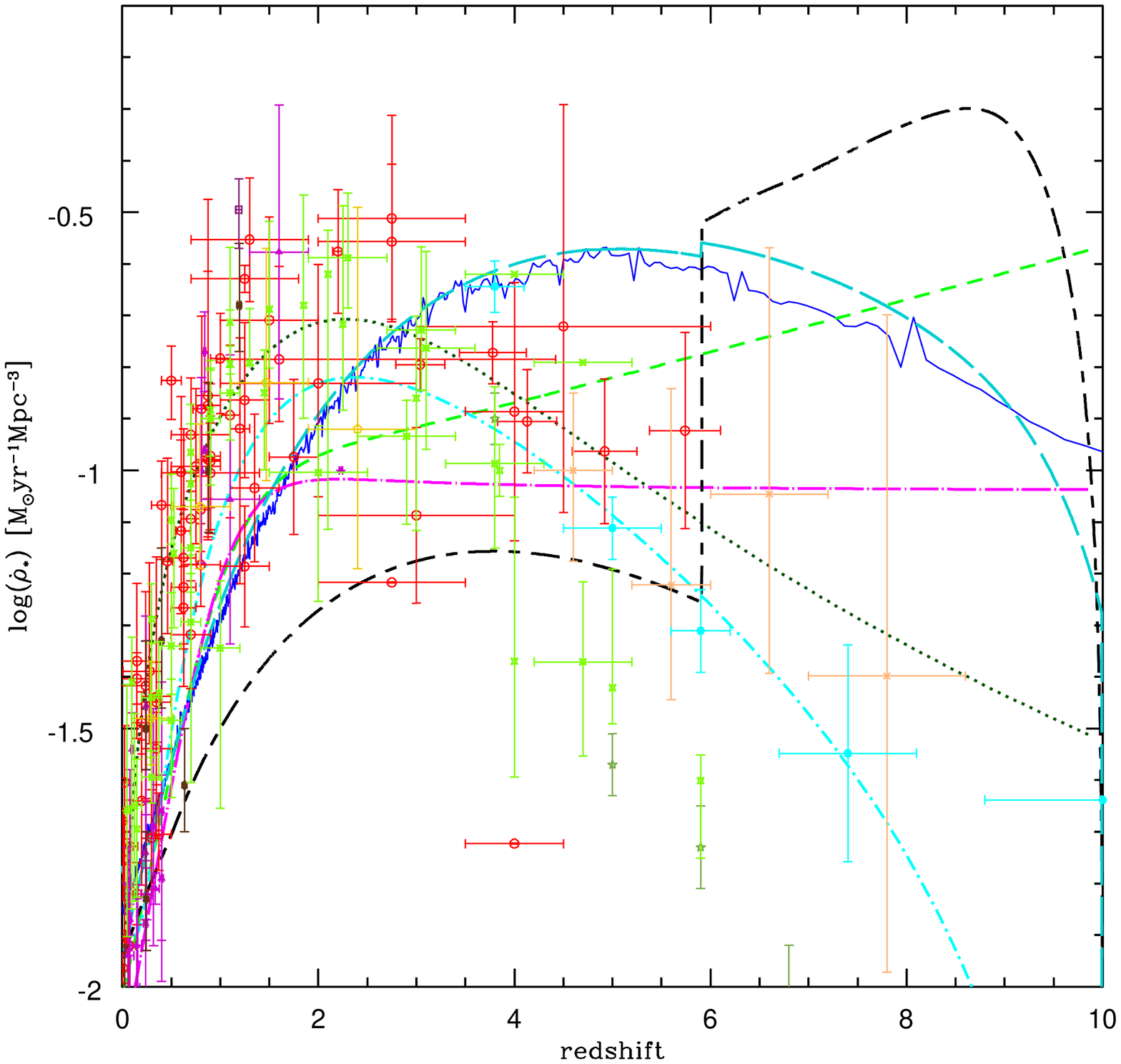}
\caption{A compilation of determinations of the CSFH, compared to the results of this work. Models are from: \citet{a45} magenta long dash-dotted line, 
        \citet{a46} green short-dashed line, \citet{a47} dark green dotted line, \citet{a15} blue solid line, \citet{a48} cyan dot short-dashed line,
        this work (PLE) black short-dashed long-dashed line, this work (NDE) turquoise long dashed line. 
        Data point are the same as Figure \ref{fig:csfr1}. }\label{fig:mcsfr}
\end{figure}

In Figure \ref{fig:mcsfr} we show the results of several determinations of the CSFR that we compare to our model predictions, in particular with 
the PLE and with our best model with NDE. The determinations of the CSFR we use for this comparison are: 
\begin{itemize}
 \item \textsl{\citet{a45}}:

  It is based on a sample of star-forming galaxies at $z\gtrsim~4$. The corresponding CSFR based on their data has been then parametrized by 
  \citet{a46} (their model SF2).
 
 \item\textsl{\citet{a47}}:

  Combining the data of the Two Micron All Sky Survey (2MASS) Extended Source Catalog and the 2dF Galaxy Redshift Survey, they produce an infrared 
  selected galaxy catalogue. They use it to estimate the  galaxy LF and to infer the total mass fraction in stars. Then they use their results together with 
  the data of \citet{a45} to obtain a parametric fit of the CSFR. Their parametric form has also been used by \citet{a57}.
 
 \item \textsl{\citet{a46}}: 
  
  Here we refer to their model SF3. It is a parameterization of the CSFR based on the data collected by the Burst and Transient Source Experiment (BATSE). 
  It is adopted to test the hypothesis that the rate of GRBs traces the global star formation history of the Universe. 
 
 \item \textsl{\citet{a48}}:

 It is a model based on a modified version of the parametric form suggested by \citet{a2}, taking into account dust extinction.
 Model parameters are determined fitting the collection of measurements of the CSFR of \citet{a58}.

 \item \textsl{\citet{a15}}:

 Prediction from a semi-analytic model for galaxy formation in the framework of the hierarchical clustering scenario.

\end{itemize}

For further information we address the reader to the above mentioned papers.

From Figure \ref{fig:mcsfr} it is possible to see that our model with number density evolution shows a complete agreement with the semi analytical 
model of \citet{a15} and, in general, shows a better agreement with the data, if compared to the determinations of \citet{a45} and \citet{a46}.
For what concerns the PLE model, this presents lower values than the other CSFRs, in the redshift range $[0-6]$. On the other hand,
for $z>6$ it exceeds all the other predictions considered here, as we have discussed already.

\subsubsection{The cosmic ISM mean metallicity}
Here we compute the evolution of the cosmic, luminosity weighted, mean interstellar metallicity of the galaxies of different 
morphological type. With this definition, we indicate the mean metallicity of the gas from which stars are born, in  a unitary volume of the Universe.
As suggested by \citet{a49}, we compute it through the following expression:
\begin{equation}
 \bar{Z}=\frac{\int{Z_{i}(L_{i})L_{i}\Phi_{i}(L_{i})\,dL_i}}{\sum_{i}{\int{L_{i}\Phi_{i}(L_{i})\,dL_i}}},
\end{equation}
where $Z_{i}(L)$ is the average interstellar metallicity in a galaxy of luminosity $L_{i}$, at any given cosmic time, and of the \textit{i}-th 
morphological type and $\Phi_{i}(L_{i})$ is the luminosity function of the \textit{i}-th morphological type.

We compute this quantity both in the case of PLE and NDE, using the metallicity evolutions obtained with the chemical evolution models, 
together with the local B-band luminosities obtained with the spectrophotometric code.
The parameters of the local luminosity functions, are from \citet{a16}, as in the previous paragraphs.
\begin{figure}
\centering 
\includegraphics[height=20 pc, width=20 pc]{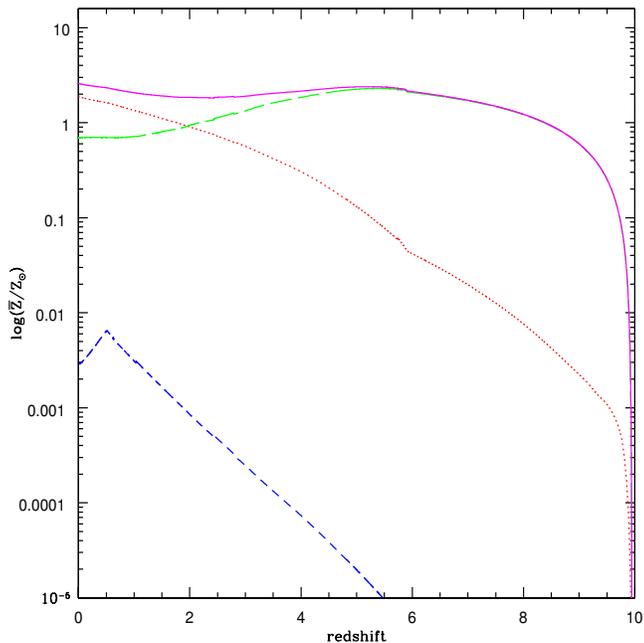}
\caption{The cosmic luminosity weighted mean metallicity relative to the Sun in the case  of pure luminosity evolution. Magenta solid line: total; 
         red dotted line: spirals; green long-dashed line: ellipticals; blue dashed line: irregulars. All the values are normalized to the solar 
         metallicty $Z_{\odot}=0.0134$, from \citet{a50}.} \label{fig:cmdple}
\end{figure}

From Figure \ref{fig:cmdple} it is clear that in the PLE scenario ellipticals dominate the mean galactic metallicity throughout the whole cosmic time.
This is due to their higher average metallicity compared to spirals and irregulars and to the constant number density of all galaxies. 
The spiral galaxies are the second most important contributors to the cosmic chemical enrichment, whereas irregulars give a negligible
contribution. We can see that our model is in agreement with \citet{a51} who found that for $z\gtrsim6$ the main metal producers are ellipticals
whereas for $z\lesssim2$ the main sources of metals production are spirals.
It is worth noting that, the decrease of the luminosity weighted metallicity of ellipticals is due to the fact that, at $z\sim6$ in our
model, ellipticals stop forming stars and from that moment on their B-band luminosity abruptly decreases. Also spirals decrease their B-band luminosity 
but mildly since their star formation lasts for the whole cosmic time. 
We note that in this case two distinct effects are involved: the metallicity which is increasing and the luminosity which is decreasing with
decreasing $z$.

\begin{figure}
\centering
\includegraphics[height=20 pc, width=20 pc]{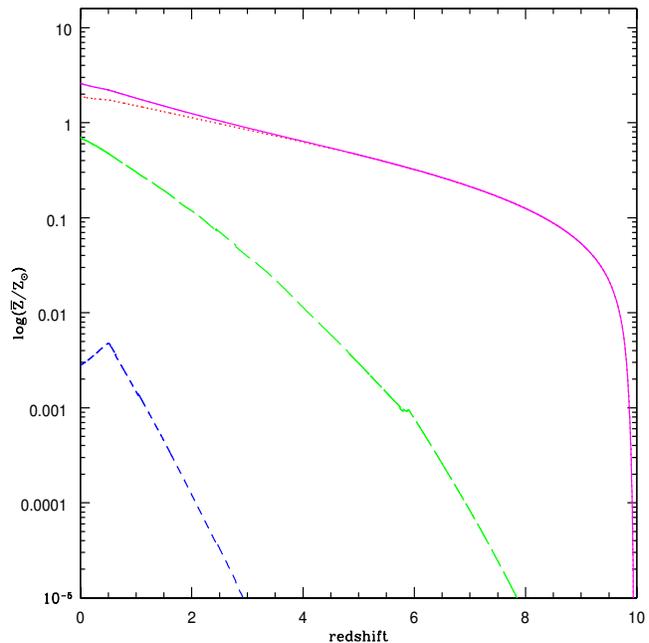}
\caption{The cosmic luminosity weighted mean metallicity relative to the Sun in the case of number density evolution. Magenta solid line: total;  
         red dotted line: spirals; green long dashed line: ellipticals; blue dashed line: irregulars. All the values are normalized to the solar 
         metallicty $Z_{\odot}=0.0134$, from \citet{a50}.}\label{fig:cmdnde}
\end{figure}

In Figure \ref{fig:cmdnde} we show the same as in Figure \ref{fig:cmdple} but under the hypothesis of number density evolution. In particular, we 
show the results of our best model, as described in section \ref{sec:nde}. The most evident fact in this case is that spirals, instead of ellipticals, 
are the main contributors to the total cosmic luminosity weighted mean metallicity. However, we need further investigation to better interpret this result. 
It has to be considered that this quantity cannot be compared directly with observations, since it is a cosmic average, while measurements are 
performed in single objects.
Another important thing we note is that, in this case the cosmic luminosity weighted mean metallicity of ellipticals is increasing throughout the 
whole cosmic time. This can be explained by the fact that the decreasing of the B-band luminosity is compensated by the increase of their number 
density at lower redshift.
We also find that our present time value of the cosmic luminosity weighted mean metallicity of the ISM in all galaxies
and in each galaxy formation scenario is $\sim2.4~Z_{\odot}$, a value that is slightly higher than $1.3~Z_{\odot}$ as found by \citet{a51}.
This is partly due to the fact that we adopt the solar metallicity of \citet{a50}, that is lower than the $Z_{\odot}$ adopted in the previous paper.

This average metallicity is not the global average metallicity of the Universe since we did not consider the metallicity of the
intracluster and intergalactic medium.

\section{Discussion and Conclusions} 
In this paper we computed the CSFR by means of detailed chemical and spectro-photometric models of galaxies of different morphological type 
(ellipticals, spirals, irregulars).These models well reproduce the main properties of local galaxies and have been tested in the past on a variety of 
observational constraints (see \citealt{a73}). Models of galactic chemical evolution of galaxies contain a few free parameters such as the efficiency of star formation, 
the slope of the IMF in the case of external galaxies, the fraction of binary systems in the IMF giving rise to SNe Ia, the timescale for gas accretion and the efficiency of 
possible galactic winds. All these parameters should be fixed by reproducing the largest as possible number of observational constraints, so that the number of well fitted 
features outnumbers the number of free parameters. For the Milky Way we have many observational features to reproduce, such as the G-dwarf metallicity distribution, the solar 
abundances, the present time amount of gas, the present time SN rates, the star formation and gas infall rates, the abundance patterns in stars ([X/Fe] vs. [Fe/H], where X 
represents the abundance of a chemical elements). Then one should reproduce the abundance gradients of chemical abundances, star formation rate and gas distribution along the disk. 
For external galaxies, such as ellipticals, the number of constraints is less than for the Milky Way but still they are enough to counterbalance the number of free parameters involved 
in the models. The observational features to reproduce are: the Color-Magnitude diagram, the average metallicity (measured by Mg and Fe) of the stars, the abundance gradients 
in the stellar populations as a function of the galactocentric distance, the high [$\alpha$/Fe] ratios measured in the dominant stellar population in the central part of these 
galaxies, the mass-metalicity relation and the [$\alpha$/Fe] vs. mass relation, the Type Ia SN rate at the present time. This last features implies that the more massive galaxies 
show higher central [$\alpha$/Fe] ratios than less massive ones. This is explained at the moment only by chemical evolution models of the knid presented here. Classical hierarchical 
models predict the contrary (see \citealt{a72}), while more recent semi-analytical models (e.g. \citealt{a67,a68}) improve the situation but they still 
do not reproduce entirely this relation. In order to obtain an increase of the [$\alpha$/Fe] ratio with galactic mass is enough to assume an increasing star formation efficiency with galactic 
mass. However, we cannot exclude a priori other solutions, such as a variable IMF from galaxy to galaxy, but this solution creates other problems (see \citealt{a69}). 
Also the models for irregulars have been tested and they reproduce the metallicity, the amount of gas, the [C,N,O/Fe] vs. [Fe/H] relations, the SN rates for an average irregular galaxy. 
Therefore, we felt reasonably confident that the star formation histories suggested by these galaxy models could be used to compute the CSFR.
We then adopted the histories of star formation predicted by such models and convolved them with the luminosity 
functions of galaxies. These LFs have been constrained to reproduce the observed local LFs.
We first computed the CSFR under the assumption of pure luminosity evolution of galaxies, namely we did assume that the main parameters of the 
LF do not vary with redshift, in particular the number density of galaxies and the slope of the LF. By doing that, 
we obtained results similar to those of \citet{a17}. Then we computed the CSFR by assuming number density and slope evolution.

Our results can be summarized as follows:

\begin{itemize}
\item In the case of pure luminosity evolution of galaxies, the CSFR traces the star formation histories of galaxies. In the framework of our galaxy 
      evolution models, ellipticals form at very high redshift in a very short interval of time and then evolve passively for the remaining galactic 
      age. The spirals and even more the irregulars, instead, are forming stars continuously and at a much lower rate for the whole galactic lifetime. 
      In this situation, we predict that the ellipticals are the dominant source of the cosmic star formation at very high redshift (from $z=6$ to $z=10$), 
      whereas spirals start to dominate for $z< 6$. On the other hand the contribution of irregulars to both the CSFR and the cosmic chemical enrichment 
      is found to be negligible. By comparing these model results with the available data for CSFR, we conclude that our models overpredict the CSFR at 
      high redshift and underpredict it at low redshift. However, most of models for CSFR in the literature also underpredict this quantity at low redshift. 
      For the high redshift situation we should take into account that the data for $z>6$ are still very uncertain, especially because of the uncertain dust 
      corrections and assumed IMF. Therefore, a conclusive statement on this point would be premature. For the low redshift we probably underestimate the amount of 
      cosmic star formation, since we did not include massive starburst galaxies at intermediate and low redshift.

\item In the case of number density evolution, by assuming that ellipticals were less at high redshift and their number increased at lower redshifts, 
      and that the contrary happened to spirals, we obtain a CSFR in much better agreement with observations. This result could imply that many 
      ellipticals formed by merging of spirals, as originally suggested by \citet{a39}.  In particular, our predicted number density evolution for 
      ellipticals implies $\phi^{*}=\phi_0 (1+z)^{-0.8}$ in agreement with \citet{a42}, whereas 
      for spirals implies  $\phi^{*}=\phi_0 (1+z)$ in agreement with \citet{a43} and \citet{a41}. Possible criticism to the scenario of the dry mergers of spirals
      to form ellipticals is given by the fact that a dry merging process seems unrealistic in the light of the difficulty of reproducing the observed local chemical 
      properties of ellipticals. In fact, it is well known that the chemical properties of spirals and ellipticals are different (e.g. different average metallicities 
      and [$\alpha$/Fe] ratios). By comparing our predicted CSFR with other determinations of the CSFR we found that it is similar to the CSFR of the hierarchical  model of Menci  \& al. (2004).

\item We tested also the case of a variation of the slope of the 
luminosity function with redshift and concluded that a variation of the slope similar to what is suggested by observations is not excluded, although it does not provide an improvement relative to 
pure luminosity evolution case and it predicts a CSFR in worse agreement with the data than the case of only galaxy number density variation. Also cases with both NDE and slope evolution at the same time 
do not provide better fits.

\item We computed the cosmic, luminosity weighted, metallicity in the gas of galaxies and its evolution with redshift. We found that, in the case of 
      pure luminosity evolution, the ellipticals dominate the cosmic metal enrichment over the whole lifetime of the Universe. On the other hand, in 
      the case of number density evolution, are the spirals to dominate the cosmic metal enrichment of the Universe.
      This is potentially an important result but it needs to be confirmed by future studies.

\end{itemize} 

\section*{Acknowledgements}
We thank F. Fontanot for many illuminating discussions. We also thank P. Tozzi and S. Vattakunnel for reading the paper and giving useful suggestions.
We thank an anonymous referee for her/his careful reading and suggestions which improved the paper.

\label{lastpage}

\end{document}